# Light-controlled multi-phase structuring of perovskite crystal enabled by thermoplasmonic metasurface

Sergey S. Kharintsev[1*], Elina I. Battalova[1], Timur A. Mukhametzyanov[2], Anatoly P. Pushkarev[3], Ivan G. Scheblykin[4], Sergey V. Makarov[3,5], Eric O. Potma[6], and Dmitry A. Fishman[6*]

[1]*Department of Optics and Nanophotonics, Institute of Physics, Kazan Federal University, Kremlevskaya, 16, Kazan, 420008, Russia*

[2]*Department of Physical Chemistry, Institute of Chemistry, Kazan Federal University, Kremlevskaya, 18, Kazan, 420008, Russia*

[3]*School of Physics and Engineering, ITMO University, St. Petersburg 197101, Russia*

[4]*Department of Chemistry, Lund University, 221 00 Lund, Sweden*

[5]*Qingdao Innovation and Development Center, Harbin Engineering University, Qingdao 266000, Shandong, China*

[6]*Department of Chemistry, University of California, Irvine, CA 92697, USA*

Halide perovskites belong to an important family of semiconducting materials with unique electronic properties that enable a myriad of applications, especially in photovoltaics and optoelectronics. Their optical properties, including photoluminescence quantum yield, are affected and notably enhanced at crystal imperfections where the symmetry is broken and the density of states increases. These lattice distortions can be introduced through structural phase transitions, allowing charge gradients to appear near the interfaces between phase structures. In this work, we demonstrate controlled multi-phase structuring in a single perovskite crystal. The concept uses cesium lead bromine (CsPbBr$_3$) placed on a thermoplasmonic TiN/Si metasurface and enables single, double and triple phase structures to form on demand above the room temperature. This approach opens up application horizons of dynamically controlled heterostructures with distinctive electronic and enhanced optical properties.

## 1. Introduction

Perovskite-structured direct bandgap semiconductors form an important class of materials with equally important applications. The presence of antibonding states near the maximum of the valence band gives rise to a defect-tolerant semiconductor material with unique electronic and optical properties, including fast charge transport, an extended free carrier diffusion length, a high exciton binding energy, and bandgap tunability.[1–6] These properties have already enabled promising applications in photovoltaics and solar energy conversion with >20% efficiency[7], as well as lasing and light-emitting devices.[3,4,6,8]

The properties of perovskite derive from its specific ABX$_3$ architecture, where A and B are cations, and X are anions arranged into chemically stable corner-sharing octahedral $BX_6$ frameworks. There are three main crystallographic phases in which perovskites exist, namely: orthorhombic ($\gamma$), tetragonal ($\beta$), and cubic ($\alpha$). Transitions between aforementioned phases contribute to the formation of multiple structural domains and lattice imperfections, specifically through crystal twinning. This latter process can be understood as an immunity response to the loss of symmetry, when, upon minimizing the Gibbs free energy, the system relaxes into a thermodynamically stable state by forming ferroelastic, near-orthogonal domains. These formations have been widely studied at the nano-, micro- and mesoscales [9–13]. In perovskites, crystal twinning is associated with lowering of the lattice symmetry by tilting and contortion of octahedrons, which causes spontaneous intrinsic stress. The introduction of such structural distortions will significantly affect local electronic structure, hence transition probabilities. Moreover, the density of states is considerably larger at these sites, where the latter can be viewed as an optical nanoantenna.[14]

Temperature[15,16] and pressure[17] perturbations have been utilized to alter the system's properties through manipulation of the crystal structure and defect density. For many perovskite systems, the bandgap for the tetragonal phase is slightly lower compared to that of the orthorhombic and cubic phases.[18,19] This causes free carriers to migrate from high to low bandgap areas, an effect that is most pronounced near phase transition sites. It has been hypothesized that such spatially non-uniform transport leads to a local build-up of free carriers in tetragonal domains.[20,21] Such an accumulation of free carriers increases the probability for electrons and holes to radiatively recombine, affecting and, ultimately, enhancing the Raman and photoluminescence efficiencies near the lattice distortions sites. *If temporally and spatially controlled, this effect could be used to actively tune the optoelectronic properties of the material, such as boosting the brightness of light-emitting devices[1,3] or modulating the lasing efficiency* [22,23]. Control of such enhanced emission requires precise manipulation of phase structuring within the single crystal. This manipulation, in turn, necessitates control of the phase transitions and thus dynamic management of the local temperature within the material.

There are two major prerequisites for producing multi-phase structures in a dynamically controlled manner – (1) a mechanism for rapid and efficient heating at the nano- to micro-scales and (2) a mechanism for heat release from the heated location. In this context, the heating mechanism at small spatial scales can benefit from the thermoplasmonic effect, through which heat can be locally generated via absorption of incident light by a plasmonically resonant structure.[24–26] This approach has been shown useful for efficient heat generation (up to a few thousands of K), followed by the rapid directional heat transfer to the material of interest.[26,27] Note that all-inorganic halide perovskites have remarkably small thermal conductivity (0.42 W m$^{-1}$K$^{-1}$),[28] yet possess high photo- and thermal stability. The latter underlines the possibility of maintaining spatial and temporal stability of the heat pattern and gradients across a single crystal, resulting in a stable multi-phase semiconducting system.

In this work, we demonstrate controlled multi-phase structuring of a single crystal of cesium lead bromine ($CsPbBr_3$). We achieve this level of control by placing the perovskite crystal on a thermoplasmonic metasurface that consists of a 2D array of stacked titanium nitride (TiN) plasmonic nano-pads on top of silicon (Si) nano-pillars (Figure 1a).[29] When irradiated with visible light at a wavelength resonant with the TiN structure, the plasmonic nano-pad serves as an *optically switchable heater*, while the Si pillar provides a channel for heat dissipation. This geometry produces sub-wavelength thermal gradients across the perovskite microplate, triggering on demand formation of stable phase domains within the original single crystal.

## 2. Results and discussion

### 2.1. Device concept

$CsPbBr_3$ perovskite undergoes two reversible phase transitions above the room temperature. These are the orthorhombic-to-tetragonal (361 K) and the tetragonal-to-cubic (403 K) phase transitions as determined with fast scanning calorimetry (FSC,

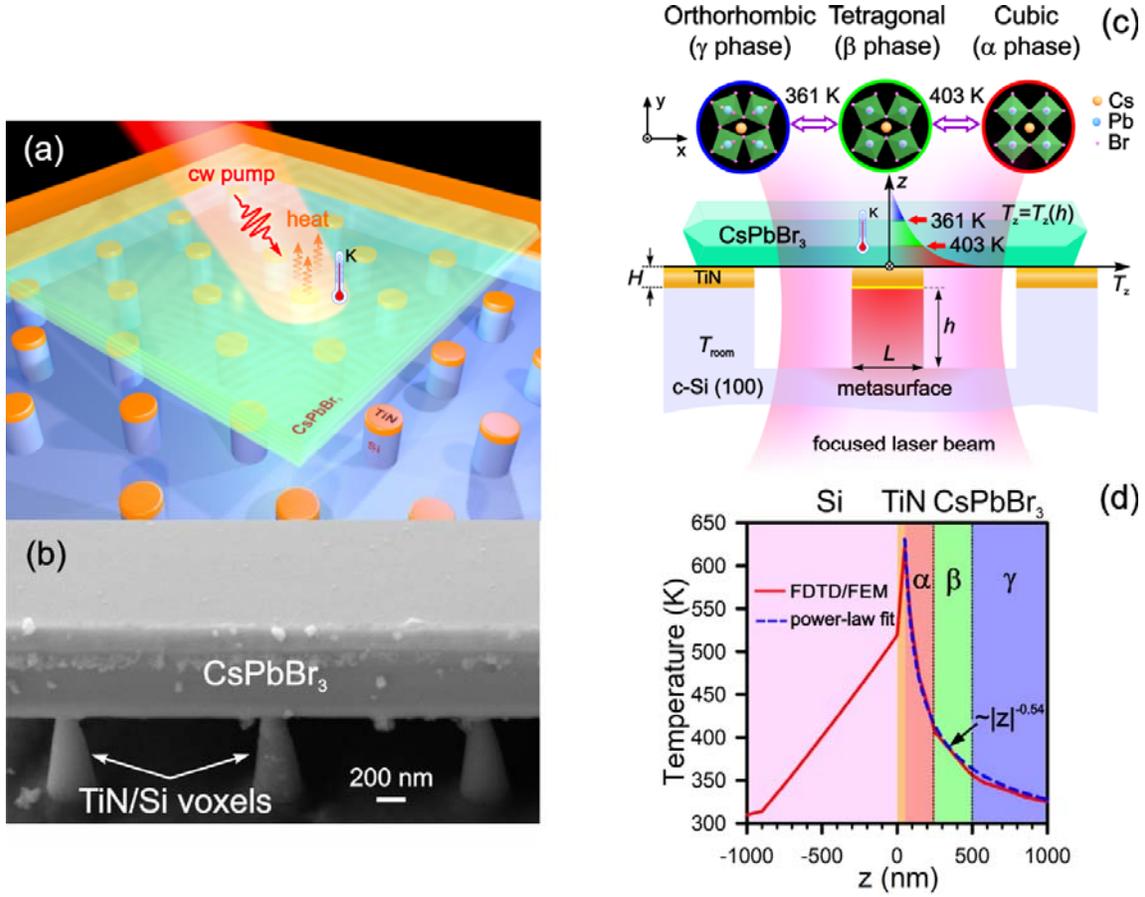

**Figure 1.** (a) Schematic representation of a $CsPbBr_3$ platelet mounted on a metasurface array. (b) A 72° tilted SEM image of the edge facet of the $CsPbBr_3$ platelet on the TiN metasurface. (c) Optical heating of halide perovskite crystal by the TiN/Si nanostructure. Color areas within the temperature gradient represent the γ phase of the Pnma space group (blue), the β phase of the P4/mbm space group (green) and the α phase of the Pm3m space group (red). (d) Finite-difference time-domain (FDTD) and finite element method (FEM) simulations of the axial temperature distribution across the TiN/Si and $CsPbBr_3$ crystal.

Supplementary Figure S1 and Supplementary Section 1). As the rate of the temperature sweep increases, the data clearly shows the lack of mirror symmetry between heating and cooling experiments. This observation points to defect-induced spatial heterogeneity within the crystal and reflects an imbalance in the potential energy barriers associated with the conversion of structure from lower to higher symmetries and *vice versa*. This makes the FSC method highly sensitive to the crystalline imperfection content and density. The presence of these phase transitions at high temperatures underlines the possibility to form a combination of different crystal phases in the material if steep and steady temperature gradients are introduced. Figure 1a schematically illustrates such a device concept that operates at ambient laboratory conditions. The $CsPbBr_3$ perovskite platelet (10 μm x 14 μm x 1 μm) is placed on a metasurface that is comprised of a hexagonal 2D array of Si pillars with a subwavelength base ($L<\lambda$). Each nanopillar is capped by a TiN plasmonic pad on top (Figures 1a and 1b). Upon illumination (continuous wave, 633 nm, 16 mW, 0.6 μm spot size, NA=0.7), the TiN pad functions as a photothermal heater, while the Si pillar transfers heat down to the bulk substrate. Silicon was chosen as the thermostat material because of its large thermal conductivity (148

W m$^{-1}$K$^{-1}$) and strong Raman response (Si-Si 521 cm$^{-1}$). Moreover, its Raman activity is temperature sensitive, permitting its use as a probe for Raman-based thermometry.

The light-to-heat conversion is expected to be maximum at the plasmonic absorption resonance of the TiN structure, as characterized by the absorption power $P = \sigma_{abs} I_0$, where $\sigma_{abs}$ is the absorption cross section and $I_0$ is the incident intensity.[24,25] The accessible temperature range at a thermal stationary state of the system will depend on several factors, namely the effective thermal conductivity of Si, the pillar's lateral and axial dimensions, the permittivity $\varepsilon$ of the TiN pad and the incident flux $I_0$. The pillar geometry, defined by base lateral size $L$ and height $h$, governs the heat dissipation efficiency and its effect has been discussed previously for composite TiN/Si rods,[29–31] tubes and trenches.[32] Taking into account the Fröhlich resonance condition, we can derive the temperature change at the top of the Si pillar as a function of structure height $h$ and incident light intensity $I_0$ as follows[29,31]:

$$\varepsilon'_{TiN}(\lambda_0) = -2\varepsilon_{Si},$$

$$\Delta T_L(h, I_0) \approx \frac{3}{4}\frac{L^2 Q^2}{\beta \lambda_0}\varepsilon''_{TiN}\left[\frac{1}{\kappa_{Si}}I_0 - \frac{\sigma_{abs}}{\kappa_{Si}^3}\frac{\partial \kappa_{Si}}{\partial T}hI_0^2\right],$$
(1)

where $\lambda_0$ is the wavelength at the plasmonic resonance, $\varepsilon_{TiN}=\varepsilon'_{TiN}+\varepsilon''_{TiN}$ is the complex permittivity of a TiN heater, $Q=-\varepsilon'_{TiN}/\varepsilon''_{TiN}$ is a Q-factor for the plasmon resonance, $\kappa_{Si}$ is the temperature-dependent thermal conductivity of bulk Si and $\beta$ is the geometry-dependent dimensionless thermal capacity of TiN.[25] For smaller pillar heights (<200 nm), the first term dominates and $\Delta T_L$ is expected to show a linear dependence on $I_0$ (Figure S2, see Supplementary Information) [29]. For taller pillars, the contribution of the second term increases accordingly, resulting in a quadratic dependence of the temperature on $I_0$. Moreover, $\Delta T_L$ is now dependent on the first temperature derivative of $\kappa_{Si}$. For bulk Si, this derivative has a negative sign above room temperature, and, thus, $\Delta T_L$ should monotonically increase with the incident intensity. For structures with height exceeding $h$>500 nm, significant deviation from experimental observations have been reported and explained in terms of thermal anisotropy.[32]

Because the thermal conductivity of Si ($\kappa_{Si} = 148$ W m$^{-1}$K$^{-1}$) significantly exceeds that both of air ($\kappa_{air}$=0.0263 W m$^{-1}$K$^{-1}$) and CsPbBr$_3$ perovskite ($\kappa_{CsPbBr_3} = 0.42$ W m$^{-1}$K$^{-1}$), the pillar structure becomes the dominant channel for heat dissipation with its geometry being the key factor in determining the steady state temperature profile. *Hence, pillars of a specific height provide access to specific temperature ranges, while fine control within this range can be realized by varying the incident light intensity $I_0$. When irradiated, an array of such nano-heaters can generate a two-dimensional temperature pattern formed by sub-wavelength hot spots ($L<\lambda$).* Induced thermal gradients along the axial direction in the perovskite allow particular phase domains to be formed, subject to the distance from the heating TiN pad (Figure 1c). Figure 1d shows a combined finite-difference time-domain (FTDT) and finite element (FEM) method (ANSYS/Lumerical) simulation of the axial temperature distribution. The simulation reveals the axial heat distribution within the layered system, comprised of a 1 μm Si pillar, a 50 nm TiN pad and a 1 μm CsPbBr$_3$ crystal. A pillar of this height is associated with a steady state temperature range of 320-520 K or a 0.23 K/nm thermal gradient within the Si material. The maximum temperature at the plasmonic structure is chosen to be 630 K, a critical temperature point beyond which the CsPbBr$_3$ optoelectronic properties drastically change.[33] The temperature gradient in the perovskite interior follows a $|z|^{-0.54}$ dependence ($R^2$=0.997), which is mainly determined by crystal thermal conductivity. In these simulations, the surrounding medium is assumed to be air. Depending on the initial nano-pad temperature (i.e. input light flux), the crystal interior can be comprised of a single γ phase or a structure of two or all three phases, as shown for $T_0$=630 K in Figure 1c.

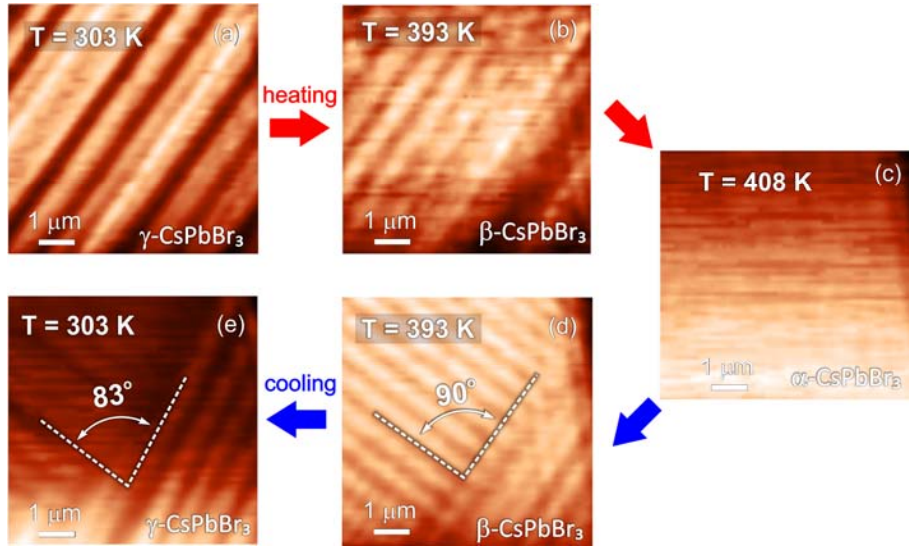

**Figure 2.** Confocal reflection images of a CsPbBr$_3$ crystal using polarized light at (a) 303 K (orthorhombic phase), (b) 393 K (tetragonal phase), (c) 408 K (cubic phase) and back (d) 393 K (tetragonal phase), (e) 303 K (orthorhombic phase).

*2.2. Optical visualization of phase transitions and crystal twinning*

The real time dynamics of domain formation and twinning in perovskite crystal on a microscale is shown in Supplementary Movie SM1. In this experiment, the crystal was placed on a hot plate to perform temperature sweeps from 340 to 410 K and back, spanning the orthorhombic-to-tetragonal-to-cubic phase transitions. The heating and cooling rates were sufficiently slow to allow a uniform temperature to establish itself throughout the crystal. Optical imaging and other spectroscopic experiments were performed with the aid of a sample piezo positioning feedback system, as described in Section 3 of Supplementary Information. This solution overcomes experimental obstacles such as the thermal expansion of the sample and setup elements. It also corrects for beam defocusing by the Bragg-like grating formed through crystal twinning within the sample volume (see *Methods* and *Supplementary Information*, Figure S3 and Section 3).

Figure 2 depicts confocal reflection images of the crystal surface at the selected steady state temperatures. At 303 K (Figure 2a), the crystal consists of parallel domains of the γ phase that are oriented at a 45° angle (<110>) relative to the lab frame. The transition to the tetragonal phase occurs around 393 K (Figure 2b). As the temperature is increased to 408 K, the stripes disappear completely, indicating the formation of a homogeneous cubic crystal (α phase, Figure 2c). As the system cools down and crosses the cubic-to-tetragonal transition, crystal twinning triggers the formation of multiple tetragonal domains (Figure 2d) and a further temperature decrease brings the crystal back into the orthorhombic phase (Figure 2e). The data shows clear differences between the images of the crystal at the same temperature points, but opposite ends of the temperature cycle (Figures 2a and 2e). This difference in the patterns further confirms the results of the FSC experiments (Figure S2), indicating that the potential energy barriers are different when the phase transition proceeds along different directions of the temperature sweep.

*The symmetries of the original and final crystallographic phases associated with a transition are the key factors in the evolution of the crystal twinning.* Upon careful examination, it is clear that the resultant orthorhombic phase reveals a $7^0$ deviation angle relative to the previously orthogonal orientation of the domains in the tetragonal phase (Figure 2d and 2e). This is in excellent agreement with previous calculations by density functional theory (DFT) that yielded a $\phi \sim 13^0$ octahedral tilt for orthorhombic

CsPbBr$_3$.[18] The rotation of the corner-sharing Br atoms of the [PbBr$_6$]$^{4-}$ octahedron in the equatorial plane by $\phi/2 \sim 6.5^0$ should result in a relative re-orientation of the domains to $90^0-\phi/2 \sim 83.5^0$, as demonstrated in Figure 3f.

*2.3. Temperature dependence of Raman and photoluminescence signatures*

Both the FSC and the optical imaging experiments reveal information about the phase transitions in the perovskite crystal, and both measurements point to the importance of lattice imperfections and distortions. To examine their role at the microscopic level, we performed Raman and photoluminescence experiments, which are particularly sensitive to the electronic structure near defects and phase interfaces.

The Raman spectrum of CsPbBr$_3$ perovskite features two main low energy vibrational modes, namely the 127 cm$^{-1}$ TO (first-order transverse optical) and the 312 cm$^{-1}$ 2LO (second-order longitudinal optical) Pb-Br stretching phonon modes (Figure 3a-d).[34] It is important to note that the presence of the 312 cm$^{-1}$ peak in the Raman spectrum is evidence for the more pristine CsPbBr$_3$ structure relative to the presence of CsPb$_2$Br$_5$, with the latter being the result of exposure to water.[35]

The temperature dependence of the Raman spectra for both the TO and 2LO modes are different for various directions of the temperature sweep (Figures 3e and 3f) of a uniformly heated crystal. As the temperature is increased, the intensity of the TO phonon line (127 cm$^{-1}$) undergoes two extrema corresponding to the orthorhombic-to-tetragonal (361 K) and tetragonal-to-cubic (403 K) phase transitions (Figure 3a and 3e). The β-CsPbBr$_3$ phase reveals an expected trend, namely the decrease of the Stokes intensity with temperature, caused by the bandgap widening and the depletion of carriers in the valence band.[18,19] Meanwhile, the temperature dependence of the Stokes bands ascribed to the α and γ phases shows the opposite trend (Figure 3e). This observation can be explained by the interplay between the thermal volumetric expansion and the tilt of the [PbBr$_6$]$^{4-}$ octahedra.[18] It has been predicted that both mechanisms are capable of significant widening of the bandgap[19], estimated to be <2.0 eV for β-CsPbBr$_3$ versus ~2.36 eV for γ-CsPbBr$_3$ and ~2.4 eV for α-CsPbBr$_3$. These bandgap variations offer possible explanations for the observed positive temperature trends. For example, for the given experiments the Raman process in β-phase is closer to the resonance for the used excitation photon energy (633 nm, 1.96 eV). This may lead to a signal increase when more β-phase sites are introduced. Another potential mechanism derives from the contribution of the shallow and deep states to the free carriers population at the conduction band is expected to increase with temperature,[36] enabling to change the Raman polarizability.[37]

When the temperature is lowered, the Raman intensity of the TO mode decreases continuously and does not exhibit any extrema in this temperature range. We speculate that such behavior can be understood from the dominant role of the crystallographic deformation of the [PbBr$_6$]$^{4-}$ backbones while cooling down. Spatially resolved Raman intensity maps for each phonon mode at different temperatures are presented in Figure S4. The images agree well with the confocal reflection images (Figure 2). They demonstrate a clear variation in the domain pattern as a function of the directionality of the temperature sweep across particular phase transitions - a result of the unequal potential energy barriers of the high-to-low and low-to-high symmetry conversions.

The difference in crystal twinning also impacts the temperature trend of the TO and 2LO phonon lines, resulting in their characteristically different behavior (Figure 3e and 3f). For the TO phonon mode, electron-phonon scattering at the Γ point is more sensitive to twin domain formation due to the overall momentum restrictions for the one-phonon process. This is opposite for the 2LO mode, for which there is a simpler path to fulfill momentum conservation due to the involvement of two phonons to scatter light inelastically. While the 2LO line clearly shows the γ→β transition (red curve, Figure 3f), at the same time it appears insensitive to the β→α transition. The cooling curve exhibits similar behavior for both transitions. Whereas the multi-phonon mode can be utilized as a temperature probe for a defect-free crystal, the single-phonon TO mode is more sensitive to the orthorhombic-to-tetragonal and tetragonal-to-cubic phase transitions.

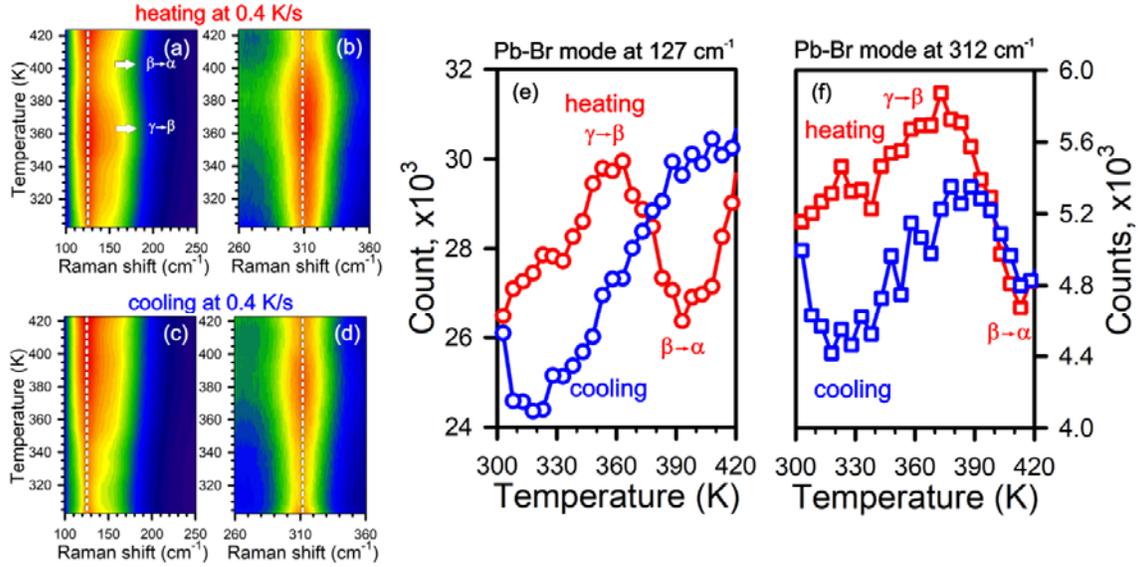

**Figure 3.** Temperature-dependent Raman spectroscopy of a CsPbBr$_3$ crystal at thermal equilibrium for TO phonon mode at 127 cm$^{-1}$ (a, c) and LO two-phonon mode at 312 cm$^{-1}$ (b, d) upon heating and cooling at a rate of 0.4 K/s. (e,f) Cross-sections at peak center for TO and 2LO modes (dashed lines in (a-d).

Photoluminescence (PL) microspectroscopy provides additional information on the carrier dynamics and the origin of the emission mechanism. The latter has been investigated through power dependence and fluorescence lifetime studies and is discussed in detail in Supplementary Information, Section 5. Here, we focus primarily on the temperature trends of perovskite photoemission. When the temperature is raised, the PL intensity drops dramatically (Figure 4a), reaching minimum at $T_{\gamma \to \beta}$ at 361 K. Further increase of the lattice temperature gives rise to a higher PL intensity for the tetragonal (β) and cubic (α) phases (red curve, Figure 4c). The overall PL spectral shape reveals complex behavior through the sweep, showing splitting-like behavior at high temperatures (Figure 4b). First, a blueshift of the mean of the spectral distribution (~16 meV) is observed (Figure 4d), indicating the bandgap expansion of the cubic phase at 423 K.[19,37] Second, a red-shifted signature (~18 meV) is observed, which is suggested to originate from the competition between surface and interior contributions of the crystals (Figure 4d).[37,38]

A radically different trend is observed when cooling is performed, with fluorescence showing a strong local maximum at the $\gamma \to \beta$ transition (Figure 4b and Figure 4c). A similar observation has been reported for methylammonium lead triiodide (CH$_3$NH$_3$PbI$_3$ or MAPbI$_3$), upon cooling from 160 K to 140 K.[20] The nature of the PL enhancement across this phase transition can be understood as resulting from the funneling effect,[20] when mobile carriers migrate to the "defect-free" low-bandgap tetragonal phase.

The observed hysteresis agrees well with the confocal reflection and Raman studies, and can be understood in a similar manner - lattice reconstruction and the dependence of crystal twinning on the sign of the temperature change $\Delta T$. Since twinning requires the base of one domain to be matched to and shared with the side of another, its probability will strongly depend on the presence of inherent crystal imperfections and the geometry of the original and resultant phases. This leads to significant differences in the overall pattern of the multi-domain assembly as a function of the sign of the temperature trend, and, in turn, the number of structural defects and phase interfaces being formed. This phenomenon also explains the striking contrast in the PL quantum efficiency. It suggests that the presence of point defects and crystal twinning favor the $\beta \rightarrow \gamma$ transition and hinder PL for the reverse direction of the transition.

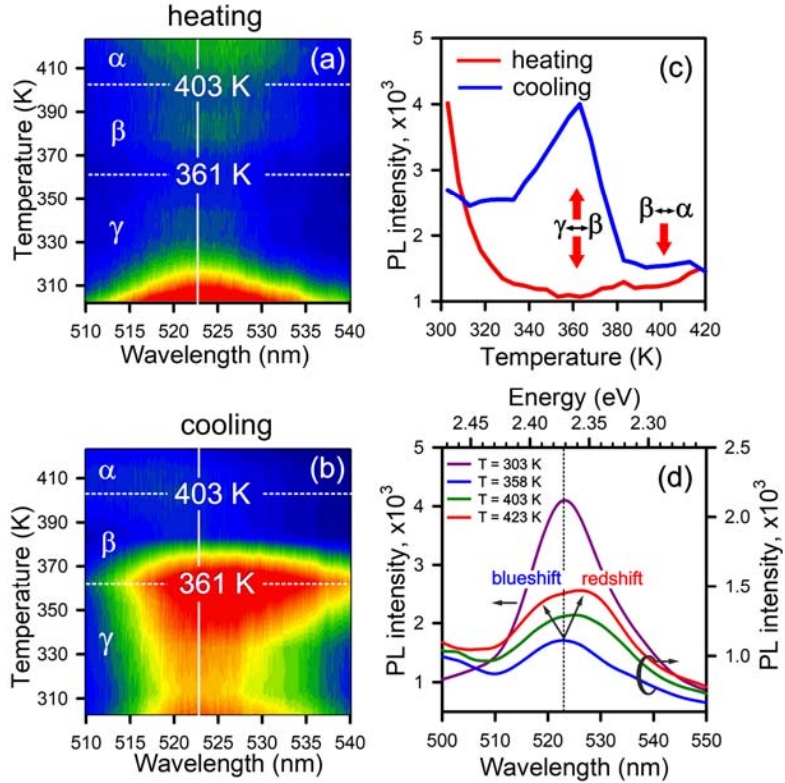

**Figure 4.** (a,b) False color PL maps for different sign of temperature sweep. (c) Cross-sections along the vertical dashed straight lines at the center of PL spectrum. (d) PL spectra for different temperatures.

*2.4. Optical properties of multiple phase single crystal.*

After the optical characterization of pervoskite platelets held at a uniform temperature, using reflection, Raman and photoluminescene microspectroscopy, we next used these optical tools to study crystals subjected to a temperature gradient. For this purpose, we employed the metasurface heating device discussed in Section 2.1 to maintain stable temperature gradients in the crystal and control the distribution of phase domains in the axial (z) direction.

Figure 5a shows a scanning electron microscopy (SEM) image of a $CsPbBr_3$ microplate placed on the metasurface. Figures 5b1 to 5b3 (cyan, yellow and magenta) depict 72° tilted images of the corners marked with an arrow of the corresponding color (Figure 5a). As is clear from the images, the structure is formed by two stacked crystal plates, most clearly observed through their exfoliation at one of the corners (Figure 5b1 and Supplementary Figure S6). The metasurface is comprised of a 2D hexagonal array of TiN/Si voxels with a pillar height estimated to be approximately 900 nm as shown in Supplementary Figure S7. The top of the voxel is visualized in the inset of Figure 5a. It is clear that, upon illumination, the TiN pads become damaged for intensities exceeding 3 MW/cm$^2$ (green and blue contoured images in the inset of Figure 5a).

The confocal reflection image at 633 nm of the perovskite platelet is shown in Figure 5c. In this image, the uncovered TiN/Si voxels have been placed at the focal plane of the objective. For such an arrangement, the voxels that are covered by the perovskite appear out of focus as light has to penetrate through the 1 μm-thick material of refractive index n=2.5.[37] This effect not only prevents efficient heating of the TiN pads, but also limits the efficient collection of the Raman signal from the voxel. The collection efficiency is instrumental, as the Raman response was utilized as a remote temperature probe. In all further experiments, the light was focused on the top of the TiN/Si voxels that are under the CsPbBr$_3$ microplate.

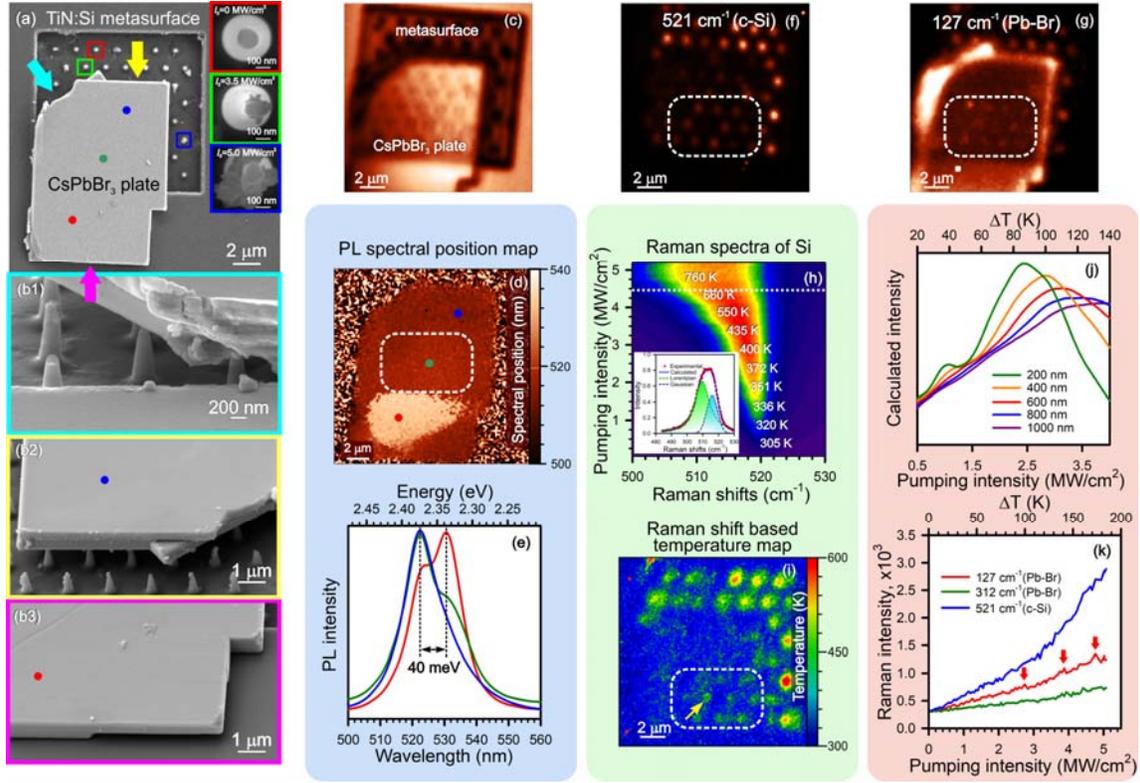

**Figure 5.** CsPbBr$_3$ platelet on the thermoplasmonic TiN/Si metasurface. (a) SEM image of the CsPbBr$_3$ plate over the metasurface. The insets show TiN/Si voxels, marked with the red, green and blue squares, exposed to 633 nm cw illumination with the intensity of 0, 3.5 and 5.0 MW/cm$^2$. (b1)-(b3) SEM images (side views at the tilt angles of 72° (b1) and 48° (b2), (b3) from the sides marked with cyan, yellow and magenta arrows in Figure 5a. (c) A confocal reflection image at 633 nm. (d) False color PL spectra central frequency map. (e) PL spectra taken at spots marked in Figure 5d with red, blue and green filled circles, respectively. (f, g) Raman maps at 521 cm$^{-1}$ (c-Si) and 127 cm$^{-1}$ (Pb-Br mode). (h) Raman spectra of Si pillar as a function of input light intensity. The inset shows a cross section along a dashed white line and numerical deconvolution of the composite band into Lorentzian and Gaussian components. (i) Temperature map measured based on Raman thermometry. (j) Simulated cumulative Raman signal from phase-structured crystal of different thicknesses. (k) Raman intensity *vs* the pumping intensity or gradient initial temperature $T_0$ temperature for 127 cm$^{-1}$ (green) and 312 cm$^{-1}$ (red) of perovskite phonon modes and 521 cm$^{-1}$ Si line.

Figure 5d displays the results of confocal PL imaging. PL spectra as a function of spectral position on the sample were collected using 1.7 W cm$^{-2}$ of 473 nm excitation. It is important to note that such low fluxes did not introduce any meaningful temperature gradients. In addition, the excitation wavelength used is far away from the absorption resonance of the plasmonic structures. The false color PL map can be divided into three characteristic regions according to PL spectral shape and central frequency position (Figure 5e) - blue (522 nm, 2.375 eV), red (531 nm, 2.335 eV) and an intermediate green region. We observe a clear correlation between the PL spectrum and the sample thickness and/or stacking. Higher energy PL, centered around 522 nm (2.375 eV), is observed in areas where two thin ~400 nm plates are stacked (blue spot in Figure 5d and Figure 5b2).

However, the spectrum is red-shifted by 40 meV at the position where the sample appears to consist of a 1 μm-thick single plate (red spot in Figure 5d and Figure 5b3). It has been suggested that the observed phenomenon is caused by the excitation of waveguide modes within the Fabry-Perot resonator through the absorption-emission-absorption mechanism [34]. If true, monitoring of the PL spectral position offers a means to probe the distribution of the perovskite thickness.

Figures 5f and 5g show confocal Raman maps for the 521 cm$^{-1}$ (c-Si peak of the pillar) and 127 cm$^{-1}$ (TO Pb-Br phonon mode of CsPbBr$_3$) lines. It is evident that not all voxels under the platelet can be clearly differentiated in the image. This is caused by the damage while positioning the perovskite on the metasurface and/or by the poor contact at certain positions. The enhanced Raman scattering of the TO mode at the crystal edges originates from structural inhomogeneities, where the density of surface states is higher (Figure 5g). For quantitative monitoring and visualization of the temperature at the voxel, we used Raman thermometry. This method, thoroughly described elsewhere[29–31] (see *Supplementary Information, Section 8*), utilizes the temperature dependent behavior of the c-Si Raman signal (521 cm$^{-1}$) as a remote probe. Through the use of an Echelle grating, the spectral resolution of the imaging system reaches 0.1 cm$^{-1}$ and enables temperature measurements with 5 K accuracy.

A detailed analysis of the open voxel temperature (blue voxel, Figure 5a) versus input flux is shown in Figure 5h. Note that the c-Si mode is asymmetrically broadened (inset, Figure 5h). This effect originates from the non-uniform heat distribution in the structure, resulting in the presence of contributions from both hot and cold portions of the material.[31] To further simplify the analysis, the spectrum was fitted with Lorentzian (hot medium contribution) and Gaussian (cold medium contribution) spectral line shapes, using a regularized least squares method ($R^2$=0.998). The intensity map in Figure 5i clearly indicates that the contribution from hot domains deviates significantly from a linear incident intensity dependence for intensities exceeding 4 MW cm$^{-2}$ (550 K). We attribute this effect to temperature dependent changes in the TiN permittivity, which affects the plasmon resonance frequency. In addition, the thermal conductivity of Si decreases when the temperature is raised.[39]

The signs of degradation of TiN pad appear at 750 K, where the Raman intensity peaks at about 5 MW cm$^{-2}$ and then shifts back to the higher energy side. This is also confirmed by previous experiments using ellipsometry on TiN films.[29] Thus, in our experiments, the incident light flux enabled access to the 293 K to 473 K temperature range – sufficient to activate all necessary structural transitions in CsPbBr$_3$ while preventing photo-damage of the plasmonic structures. Figure 5h shows the resulting temperature map derived from the Raman shift using Equation S1 (see Figure S9) and measured at 3.5 MW cm$^{-2}$.

The local generation of hot spots produced thermal gradients throughout the perovskite crystal. This effect resulted in the simultaneous formation of multiple phase domains, as illustrated in Figure 1. However, there are significant and fundamental differences between the temperature trends of the Raman signals when (1) heating of the whole crystal by a hot plate to achieve a uniform temperature profile, as opposed to (2) establishing a temperature gradient in the crystal with the metasurface. For the first case, the upward temperature trend discussed in *Section 2.3* is shown by the red curve in Figures 3e. This profile shows clear extrema at phase transitions with an overall signal intensity decrease across the tetragonal phase. In the second case, when the crystal is locally heated by the plasmonic structures, the temperature gradient induces multiple phases in the axial direction. For this case, *the Raman response is the cumulative signal from all the phases in the collection volume.*

The trend of the cumulative Raman signal $I_R$ *versus* the plasmonic pad temperature $T_m(0)$ should directly reflect the process of multi-phase structuring of the perovskite. The trend can be modeled as discussed in Supplementary Information, Section 10. For simplicity, one can assume one-dimensional heat dissipation in a homogeneous perovskite crystal, in which the temperature profile obeys a $T_m(z) = T_m(0) |z|^{-0.54}$ power law in the axial direction (Figure S10). The resulting Raman response can then written be as:

$$I_R = \int_0^{\Delta z} \langle I(z) \rangle dz \quad (2),$$

where $\langle I(z) \rangle$ is the average Raman signal of the homogeneous media at a given z-plane (Figure S10)

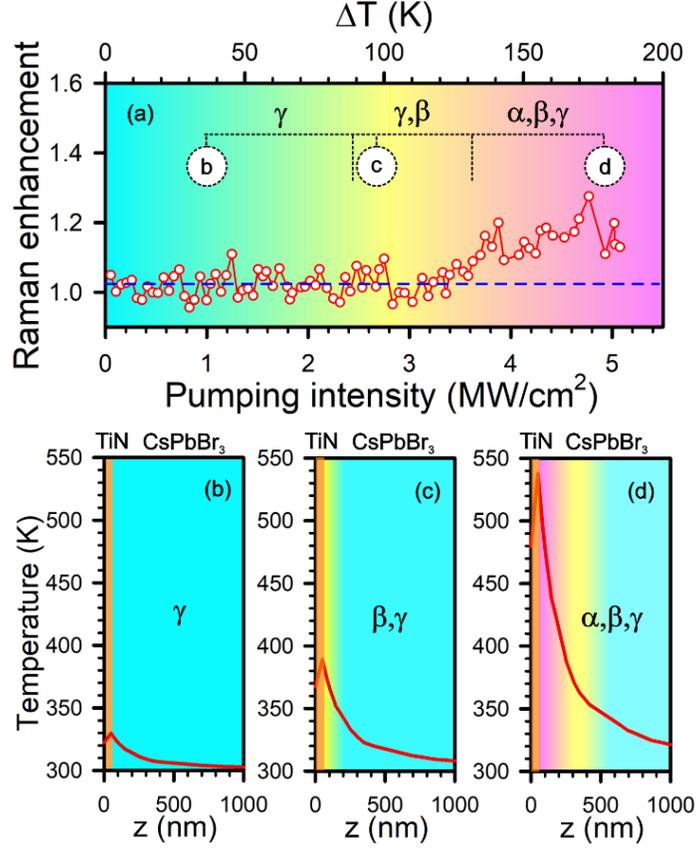

**Figure 6.** (a) Linearly corrected Raman response *versus* the incident light flux. (b, c, d) represents simulated temperature points where single, double and triple phase structure occur as seen in Raman signatures from (a). The blue dashed line represents the subtracted linear contribution.

and $\Delta z$ is the total crystal thickness. Figure 5j shows plots of $I_R$ vs $I_0$ for different $\Delta z$ of perovskite. As expected, for very thin crystals (<200 nm), the temperature trend of Raman signal closely follows the one previously observed for a thermally equilibrated crystal on a hot plate (green curve Figure 5j, red curve Figure 3e). For thicker crystals, multiple phases can contribute to the observed Raman signal. The local maximum remains highly pronounced over the monotonically increasing Raman response, indicating the formation and growth of a two-phase structure (γ and β) in the axial direction.

This model agrees well with experimental observations. Figure 5k shows the intensity evolution of the phonon modes (TO 127 cm$^{-1}$, 2LO 312 cm$^{-1}$) along with the c-Si line (521 cm$^{-1}$) as a function of the incident light flux/pad temperature. As expected, the trends for the 127 cm$^{-1}$ (Pb-Br) and 312 cm$^{-1}$ (Pb-Br) modes are inherently different from the case of the thermally equilibrated system (red and green curves, Figures 5k). For the TO mode, a clear presence of local maximum around the temperature of γ-to-β transition is observed, indicating the formation of a two-phase structure. Upon further increase of the pad temperature $T_m(0)$, another shallow bump at $\Delta T \sim$ 140 K indicates triple phase formation (α, β and γ).

To visually emphasize these signatures, the linear contribution to the Raman-temperature trend has been subtracted in Figure 6a. The linear contribution has been determined from a simple linear fit over 0 – 2.5 MW cm$^{-2}$ range (Figure 6a point b, Figure 6b), where the whole perovskite crystal remains in the single orthogonal γ phase and the intensity of the TO Raman peak should linearly increase with the incident excitation flux and temperature (Figure 3e). Upon increasing the incident intensity, the formation of the tetragonal phase at the interface of the perovskite and TiN is expected to occur at point c (Figure 6a). At this point of the trend, the steep temperature gradient creates a spatially sharp defect area where the crystal is in a transitional form between the orthorhombic and tetragonal phase (Figure 6c). This scenario manifests itself as a shallow Raman intensity maximum. Further increase of the plasmonic pad temperature drives the β-phase deeper into the crystal bulk. At $\Delta T$~130 K another shallow maximum of the TO Raman peak indicates the formation of the α-phase in close proximity to the TiN structure. Upon subsequent increase of the incident light flux, the α and β phases extend further into the crystal and significantly broaden (Figure 6d). This is expected to result in the smearing of the boundaries between the different phases. This effect is spatially asymmetric, i.e. different for the left (α-β) and the right (β-γ) sides of β phase, following the highly nonlinear temperature gradient. At higher temperatures, the signal is an interplay of several contributions, in particular the phase layer thickness, the temperature, position along the gradient, and the sharpness of the phase boundary. We hypothesize the third maximum at $\Delta T = 170$ K is the result of such a cumulative effect and may be associated with the delocalized (disordered or randomly located) phase boundary. Among other instrumental contributors to the spatial phase formation are the crystal intrinsic defects. Their presence can trigger the spontaneous formation of different phases, resulting in a highly irregular structural front. At higher TiN temperatures, the α-β and β-γ boundaries broaden significantly and may capture more defects into these areas where the phases are highly mixed.

These experiments demonstrate that the Raman signal from perovskite subjected to a stable temperature gradient, as shown in Figure 5k and emphasized in Figure 6, reveals distinct behavior that is in contrast to a bulk crystal held at uniform temperature. The dependence of the Raman signal on the incident intensity shows clear signatures of particular phase formations, their extension into the bulk of the material, and, overall, the multi-phase structuring process. Moreover, it shows that the detected Raman signal exhibits notable gain across $\Delta T$=150 K. Using Raman microspectroscopy as a probe, these results indicate that it is possible to generate on demand a distribution of single, double and triple phase structures in perovskite by simply controlling the incident light intensity.

## 3. Conclusions

In this work, we have demonstrated the proof-of-principle multiphase structured single crystal $CsPbBr_3$ halide perovskite. We have shown, the single, double and triple phase systems can be created in optically controlled fashion on the thermoplasmonic metasurface using the continuous wave illumination of modest intensities. Light-induced heat from plasmonic TiN nanopads forms strong temperature gradients within crystal bulk that are followed by sequence of corresponding phase transitions.

Lattice distortions, defects and impurities operate like an optical nanoantenna, increasing the density of states. Thus, multi-phase perovskite structures hold many interesting properties and open exciting possibilities. In such system, charge carriers migrate from lower symmetry lattice with large bandgap (orthorhombic and cubic) to higher symmetry, but lower bandgap crystal parts (tetragonal). There, highly concentrated and in close proximity to the boundaries, carriers efficiently recombine, leading to areas with significant enhancement of the optical emission. This multi-structured system promises to be highly beneficial to the development of next-generation ultracompact broadband light-emitting diodes showing high PL quantum yields above room temperatures.

## Methods

*Synthesis of CsPbBr₃ structures*

Perovskite microcrystals on glass substrates were synthesized by using a protocol similar to the previously reported.[40] $PbBr_2$ (110 mg) and CsBr (62 mg) were mixed and dissolved in 3 ml of anhydrous dimethyl sulfoxide (DMSO) inside a nitrogen-filled glovebox. Droplet of the prepared solution (volume 2 µl) was drop-casted on the substrate at ambient conditions. After that, the substrate was sealed in a preheated up to 60 ºC Petri dish containing 200 µl of liquid mixture. The solution was dried in the presence of azeotropic vapor for 5 min. As a result, the randomly oriented separate CsPbBr3 microcrystals were formed on the substrate.

*Synthesis, nanofabrication and characterization of a TiN/Si metasurface*

TiN thin films on c-Si (100) substrates were DC magnetron sputtered from a Ti target in the Ar/N$_2$ environment with a volume proportion of 30:70 at elevated temperature of 350 ºC and base pressure of $3 \cdot 10^{-9}$ mbar and power of 200 W. Prior to the film growth, the c-Si substrate was sonicated in acetone for 15 min. The thickness of the TiN films, equal to $50 \pm 5$ nm, was measured with a contact profilometer Alpha Step 200.

A 2D array of TiN/Si voxels were engraved with the help of focused ion beam (FIB) milling at a lower current of 1 pA by using Quanta 3D FEG (FEI). Since the higher TiN/Si voxels are exposed to FIB for a longer time, their lateral size is reduced due to edge melting. To avoid this detrimental effect, we used different mask templates for short and long voxels so that their lateral size is the same regardless of height.

The temperature-dependent permittivity of TiN thin films were measured with a spectroscopic ellipsometer (VASE, J. A. Woollam) within the spectral range of 250-2500 nm. The incident angle was 70°. The TiN sample was exposed to thermal annealing at the fixed temperature, whereas its permittivity was probed at room temperature. The temperature increment for each subsequent cycle was 100°C. The temperature ranged from 25 ºC to 600 ºC. The samples were annealed at ambient air for 30 min using a heating stage (Linkam Scientific Model THMS600). The heating and cooling rates were 150 °C/min and 100 °C/min, respectively.

*Fast Scanning Calorimetry*

The Fast Scanning Calorimetry (FSC) curves were registered on FlashDSC2+ (Mettler-Toledo, Greifensee, Switzerland) equipped with TC100MT intracooler with UFH1 sensor. The temperature calibration was performed using biphenyl ($T_m = 69.2$ ºC) and benzoic acid ($T_m = 122.3$ ºC) as standards to ±1 °C.

A single perovskite crystal was placed in the center of the calorimetric sensor. To improve signal-to-noise ratio the crystal size was chosen such as to almost match the active area of the sensor. Within the temperature range from 20 °C to 180 °C the sample was chemically stable, and the curves were repeatable, which allowed for averaging multiple scans to further improve signal-to-noise ratio.

*Atomic force microscopy*

The multimode scanning probe microscope NTEGRA PRIMA (NT-MDT) was utilized for visualizing a topography of the CsPbBr$_3$ microplate surface and the thermoplasmonic metasurface. The AFM probes of the "VIT_P" series with resonant frequencies around 350 kHz were used in AFM measurements. The CsPbBr$_3$ microplate mounted on the metasurface fabricated by focused ion beam milling was measured in tapping mode with a free amplitude $A_0$ of 10-20 nm and a set-point value of $A_0/2$.

*Far- and near-field Raman spectroscopy and microscopy*

Raman spectra and maps were captured with a multi-purpose analytical instrument NTEGRA SPECTRA™ (NT-MDT) in inverted configuration. The confocal spectrometer was wavelength calibrated with a crystalline silicon (100) wafer by registering the first-order Raman band at 521 cm$^{-1}$. A sensitivity of the spectrometer was as high as ca. 3000 photon counts per 0.1 s provided that we used a 100× objective (N.A.=0.7), an exit slit (pinhole) of 100 μm and a linearly polarized light with the wavelength of 632.8 nm and the power at the sample of 16 mW. No signal amplification regimes of a Newton EMCCD camera (ANDOR) was used.

128x128 pixel Raman maps were raster scanned with an exposure time per pixel of 0.1 s and were finally collected with the EMCCD camera cooled down to -95ºC. Raman spectra within the range of from -2000 to 2000 cm$^{-1}$ were registered with a spectral resolution of 0.1 cm$^{-1}$ using the Echelle grating.

*Fluorescence Lifetime Imaging Microscopy*

To measure the PL decay time we used a system built-in the confocal optical spectrometer (NTEGRA SPECTRA) that includes a picosecond diode laser (BDL-SMN) generatig pulses of 473 nm wavelength, 30 ps pulse duration, and 80, 50, or 20 MHz repetition rate, a Simple-Tau 150 TCSPC FLIM module (Becker&Hickl), and a HPM-100-40 GaAsP hybrid detector (Becker&Hickl). The detector has a detection efficiency of about 50% and is free of afterpulsing.

*FDTD/FEM calculation*

3D simulation of optical absorption of a TiN/Si voxels consisting of stacked TiN and Si cylinders under cw illumination was performed by using an Ansys/Lumerical FDTD solver. The height of the TiN pad was 50 nm, whereas the height of the Si pillar 900 nm. To avoid anomalous electric fields near the TiN pad edge we used disks with rounded edges (10 nm rounding). A mesh overlayer of 1 nm was utilized around the TiN pad and a rougher 10 nm mesh for the rest of the structure. Perfectly matching layers were used as boundary conditions for three directions. The optical and thermal properties of Si and air were imported from the Ansys/Lumerical material database. The TiN pad was exposed to a 632.8 nm focused laser light (NA= 0.7) with the intensity of 5 MW/cm$^2$. The temperature profile was calculated through an Ansys/Lumerical FEM solver in the steady state regime. The thermal conductivity of all constituents is assumed to be temperature-independent. The boundary condition of $T = 300$ K was set at the $z_{min} = -20000$ nm of the 20×20×5 μm$^3$ simulation region.

**Conflicts of interests**
There are no conflicts to declare.

**Acknowledgement**
This work was supported by grant No. 19-12-00066-P of the RSF. The PL decay time measurements were granted by the Kazan Federal University Strategic Academic Leadership Program (PRIORITY-2030). The authors acknowledge a technical support from our industrial partners: SCANSENS (GmbH, Germany) and NT-MDT BV (The Netherlands).